\documentclass[prl,superscriptaddress,preprint,endfloats,nopacs]{revtex4}

\usepackage{amsmath, amsthm, amssymb, pifont, wasysym}

\usepackage{graphicx}
\usepackage{bm} 
\usepackage{pifont}
\usepackage{dcolumn} 
\usepackage{color}
\usepackage{mhchem}
\usepackage{gensymb}
\usepackage{xcolor}
\definecolor{mygray}{gray}{0.5}
\begin{document}

\title{Stark difference in the in-plane anomalous Hall response \\in Zintl compounds \ce{Eu$A$2Sb2} (\ce{$A$} = \ce{Zn}, \ce{Cd}) thin films}

\author{Hsiang Lee}
\affiliation{Department of Physics, Institute of Science Tokyo, Tokyo 152-8551, Japan}
\author{Shinichi Nishihaya}
\affiliation{Department of Physics, Institute of Science Tokyo, Tokyo 152-8551, Japan}
\author{Markus Kriener}
\affiliation{RIKEN Center for Emergent Matter Science (CEMS), Wako 351-0198, Japan}
\author{Jun Fujioka}
\affiliation{Department of Materials Science, University of Tsukuba, Tsukuba 305-8577, Japan}
\author{Ayano Nakamura}
\affiliation{Department of Physics, Institute of Science Tokyo, Tokyo 152-8551, Japan}
\author{Yuto Watanabe}
\affiliation{Department of Physics, Institute of Science Tokyo, Tokyo 152-8551, Japan}
\author{Hiroaki Ishizuka}
\affiliation{Department of Physics, Institute of Science Tokyo, Tokyo 152-8551, Japan}
\author{Masaki Uchida}
\email[Author to whom correspondence should be addressed: ]{m.uchida@phys.sci.isct.ac.jp}
\affiliation{Department of Physics, Institute of Science Tokyo, Tokyo 152-8551, Japan}

\begin{abstract}
Recent observation of the in-plane anomalous Hall effect in magnetic Weyl semimetal \ce{EuCd2Sb2} has drawn attention to out-of-plane orbital magnetization induced by an in-plane field component. Here we study \ce{EuZn2Sb2}, a sister compound of \ce{EuCd2Sb2}, to demonstrate sensitive changes of the in-plane anomalous Hall effect on the band modulation. The Hall resistivity measured with rotating the magnetic field within the (001) principal plane of \ce{EuZn2Sb2} films exhibits a clear three-fold component corresponding to the in-plane anomalous Hall effect, which is distinct from the two-fold component of the planar Hall effect. The in-plane anomalous Hall effect of \ce{EuZn2Sb2} is highly contrasting to \ce{EuCd2Sb2}, especially in terms of its opposite sign and field dependence, which can be explained by model calculations with different band inversion parameters. Our results pave the way for systematically controlling the in-plane anomalous Hall effect and orbital magnetization through elaborate band engineering.
\end{abstract}

\maketitle
\newpage

The Berry curvature is a local quantum geometric property of electron wave functions in the band structure, underlying many physical phenomena. Among them, the anomalous Hall effect (AHE) is one of the most representative examples of macroscopic response due to the Berry curvature in time-reversal symmetry broken systems \cite{nagaosa2010anomalous,fang2003anomalous,yao2004first}. For most observed AHE, the Hall vector is parallel to the direction of the magnetic field. On the other hand, the in-plane anomalous Hall effect (iAHE), with an out-of-plane Hall vector component induced by an in-plane magnetic field, is also allowed under certain symmetry conditions \cite{nakamura2024observation,peng2024observation,liu2013plane,ren2016quantum,zhang2019plane,ren2020engineering,sun2022possible,li2022chern,li2023planar,cao2023plane,liang2018anomalous,zhou2022heterodimensional,wang2024observation,zyuzin2020plane,wang2024orbital,battilomo2021anomalous,cullen2021generating,roman2009orientation,li2023switchable}. Following reports in \ce{ZrTe5} \cite{liang2018anomalous} and \ce{VS2}/\ce{VS} \cite{zhou2022heterodimensional}, iAHE has been recently observed in several magnetic materials, including \ce{Fe3Sn2} \cite{wang2024observation}, \ce{EuCd2Sb2} \cite{nakamura2024observation}, and \ce{Fe} \cite{peng2024observation}, uncovering a new aspect of magnetotransport. On the other hand, different mechanisms of iAHE have been suggested for each material, including an anomalous orbital polarizability model \cite{wang2024orbital}, an orbital magnetization model \cite{nakamura2024observation}, and an octupole model \cite{peng2024observation}. 

To observe iAHE arising only from the orbital magnetization, Eu-based Zintl compounds \ce{Eu$A$2$X$2} (\ce{$A$} = \ce{Zn}, \ce{Cd}, \ce{$X$} = \ce{P}, \ce{As}, \ce{Sb}) with a trigonal $D_{\text{3}d}$ lattice can be seen as a highly symmetric system suitable for studying iAHE on the principal plane, some of the \ce{Eu$A$2$X$2} compounds are seen as Weyl semimetals or Weyl semimetal candidates with broken time-reversal symmetry in the forced ferromagnetic state \cite{wang2019single,su2020magnetic,ohno2022maximizing,nakamura2024berry,sprague2024observation}. In the magnetic Weyl semimetal \ce{EuCd2Sb2}, the observation of iAHE has been explained by field-induced out-of-plane Weyl points splitting and orbital magnetization \cite{nakamura2024observation}. While \ce{EuCd2Sb2} has been predicted to own four to five pairs of Weyl points \cite{su2020magnetic,ohno2022maximizing}, in \ce{EuZn2Sb2} the whole \ce{Cd} atoms are replaced with \ce{Zn}, which effectively weakens the SOC. Previous studies have suggested that \ce{EuZn2Sb2} hosts a pair of Weyl points above the Fermi level \cite{sprague2024observation}. Also, \ce{EuZn2Sb2} shows the same in-plane A-type antiferromagnetic ordering as in \ce{EuCd2Sb2} \cite{weber2006low, zhang2008new,singh2024large}. 

Here we report systematic transport measurements on single-domain \ce{EuZn2Sb2} thin films grown by molecular beam epitaxy (MBE). Similar to \ce{EuCd2Sb2} \cite{ohno2022maximizing,nakamura2024berry}, \ce{EuZn2Sb2} exhibits a large out-of-plane AHE, and its Hall angle increases upon a decrease in hole density. On the other hand, a three-fold iAHE response with respect to the in-plane field rotation emerges with an opposite sign and different field dependence in stark contrast to \ce{EuCd2Sb2}. The result can be qualitatively explained by model calculations incorporating the lifting of the band inversion.

Single-crystalline \ce{EuZn2Sb2} (001) thin films were grown on \ce{CdTe} (111)A substrates by the MBE technique following similar growth processes previously reported for other \ce{Eu$M$2$X$2} films \cite{su2020magnetic,ohno2022maximizing,nishihaya2024intrinsic}. To remove oxides on the \ce{CdTe} substrates, \ce{CdTe} (111)A substrates were first etched by 0.01\% \ce{Br2}-methanol, and then annealed around 550 \degree C under Cd flux before the \ce{EuZn2Sb2} growth. The growth temperature of the \ce{EuZn2Sb2} films was about 265 \degree C. The flux ratio of the three elements, Eu, Zn, and Sb, during the growth, was $P_{\text{Eu}}:P_{\text{Zn}}:P_{\text{Sb}}=1:7-15:2-3$. We grew about 50 nm thick films with a wide range of hole densities from $10^{19}$ to $10^{20}$ cm$^{-3}$ by adjusting the flux ratio.
Transport measurements were performed by making Hall bar devices from \ce{EuZn2Sb2} thin films and adopting a conventional 4-probe method. Low-temperature magnetotransport was measured using a low-frequency lock-in technique in a Cryomagnetics cryostat system equipped with a superconducting magnet. Magnetization curves were measured using a superconducting quantum inference magnetometer in a Quantum Design Magnetic Property Measurement System. We measured reflectivity spectra at room temperature using a Fourier transform spectrometer (the grating-type monochromator) in the photon energy region of 0.06 - 0.9 eV (0.7 - 4.6 eV), and then the optical conductivity spectra were obtained by Kramers-Kronig analysis. 

\begin{figure}
\begin{center}
\includegraphics[width=17.5cm]{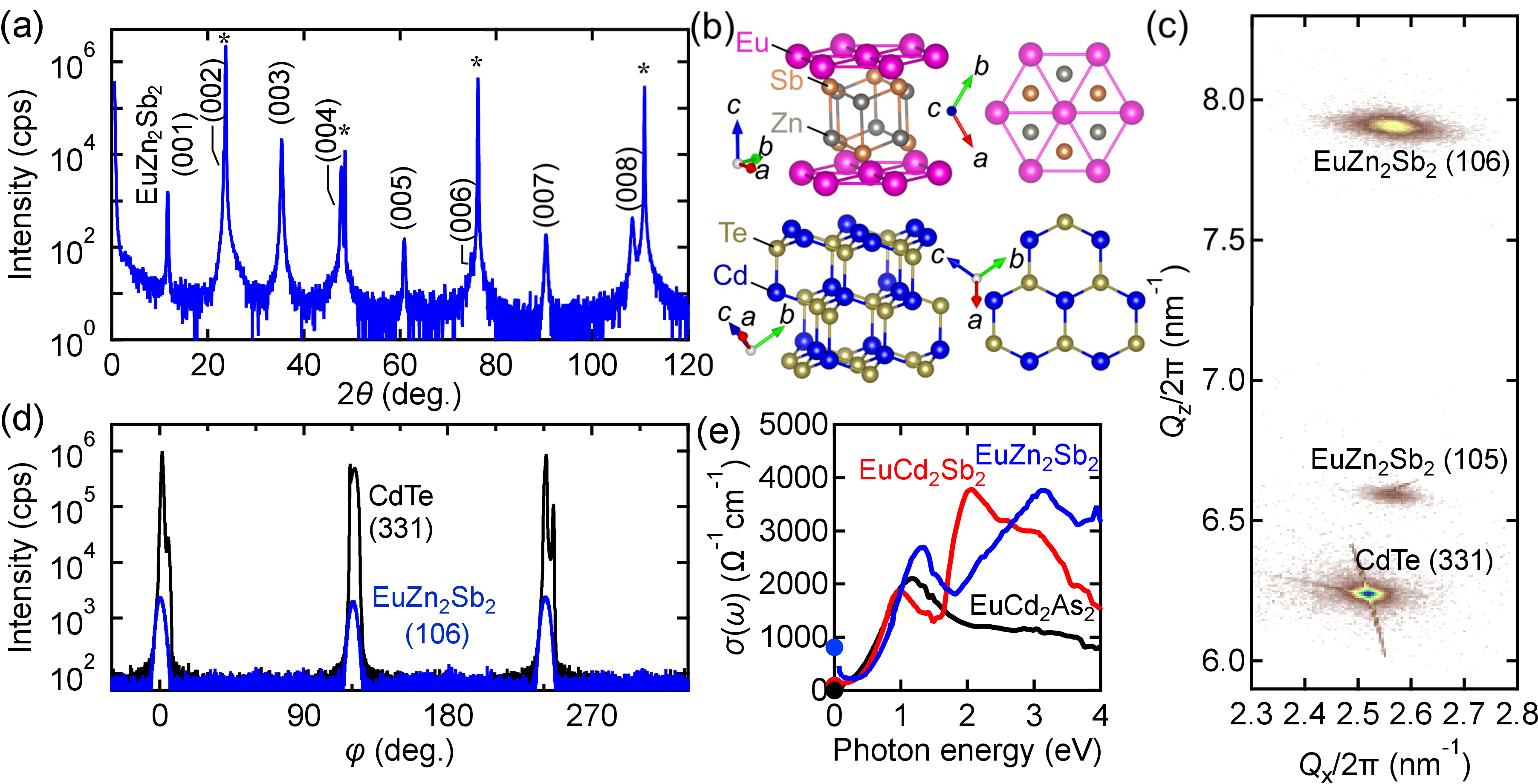}%
\caption{Structural and optical characterization of \ce{EuZn2Sb2} thin films. 
(a) XRD $\theta$-2$\theta$ scan of a \ce{EuZn2Sb2} thin film grown on a \ce{CdTe} substrate.
Substrate peaks are marked with an asterisk.
(b) Epitaxial relation between \ce{EuZn2Sb2} and the \ce{CdTe} substrate.
(c) Reciprocal space map around \ce{CdTe} (331), \ce{EuZn2Sb2} (105) and \ce{EuZn2Sb2} (106) peaks.
(d) In-plane $\varphi$-scans of \ce{CdTe} (331) and \ce{EuZn2Sb2} (106) Bragg peaks.
(e) Optical conductivity spectra of \ce{EuZn2Sb2}, \ce{EuCd2Sb2}, and \ce{EuCd2As2} films measured at room temperature.
DC conductivities are indicated by a closed circle.}
\label{fig1}
\end{center}
\end{figure}

Figure \ref{fig1}(a) shows x-ray diffraction (XRD) $\theta$-2$\theta$ scan of a \ce{EuZn2Sb2} thin film. Clear peaks of the \ce{EuZn2Sb2} \{001\} series are observed along with the \ce{CdTe} \{111\} series. To confirm the epitaxial relation as shown in Fig. \ref{fig1}(b), we performed reciprocal mapping including \ce{CdTe} (331) and \ce{EuZn2Sb2} (105) and (106) Bragg peaks. The map in Fig. \ref{fig1}(c) indicates that the [100] \ce{EuZn2Sb2} film direction aligns with the [11$\overline{2}$] direction of the \ce{CdTe} substrate. The $\varphi$-scan of \ce{CdTe} (331) and \ce{EuZn2Sb2} (106) peaks in Fig. \ref{fig1}(d) ensures that the present \ce{EuZn2Sb2} film has only a single domain without 60-degrees rotated domains. In Fig. \ref{fig1}(e), optical conductivity spectra obtained for the \ce{EuZn2Sb2} film are compared to films of two other \ce{Eu$A$2$X$2} compounds, \ce{EuCd2Sb2} and \ce{EuCd2As2}. The DC conductivity obtained from the transport measurement matches the extrapolation of the optical conductivity spectra to 0 eV. A peak at about 1 eV shifts toward higher energies in \ce{EuZn2Sb2}, suggesting that the band inversion energy in \ce{EuZn2Sb2} is smaller than in \ce{EuCd2Sb2} and \ce{EuCd2As2}, reflecting the weaker SOC.

\begin{figure}
\begin{center}
\includegraphics[width=13.0cm]{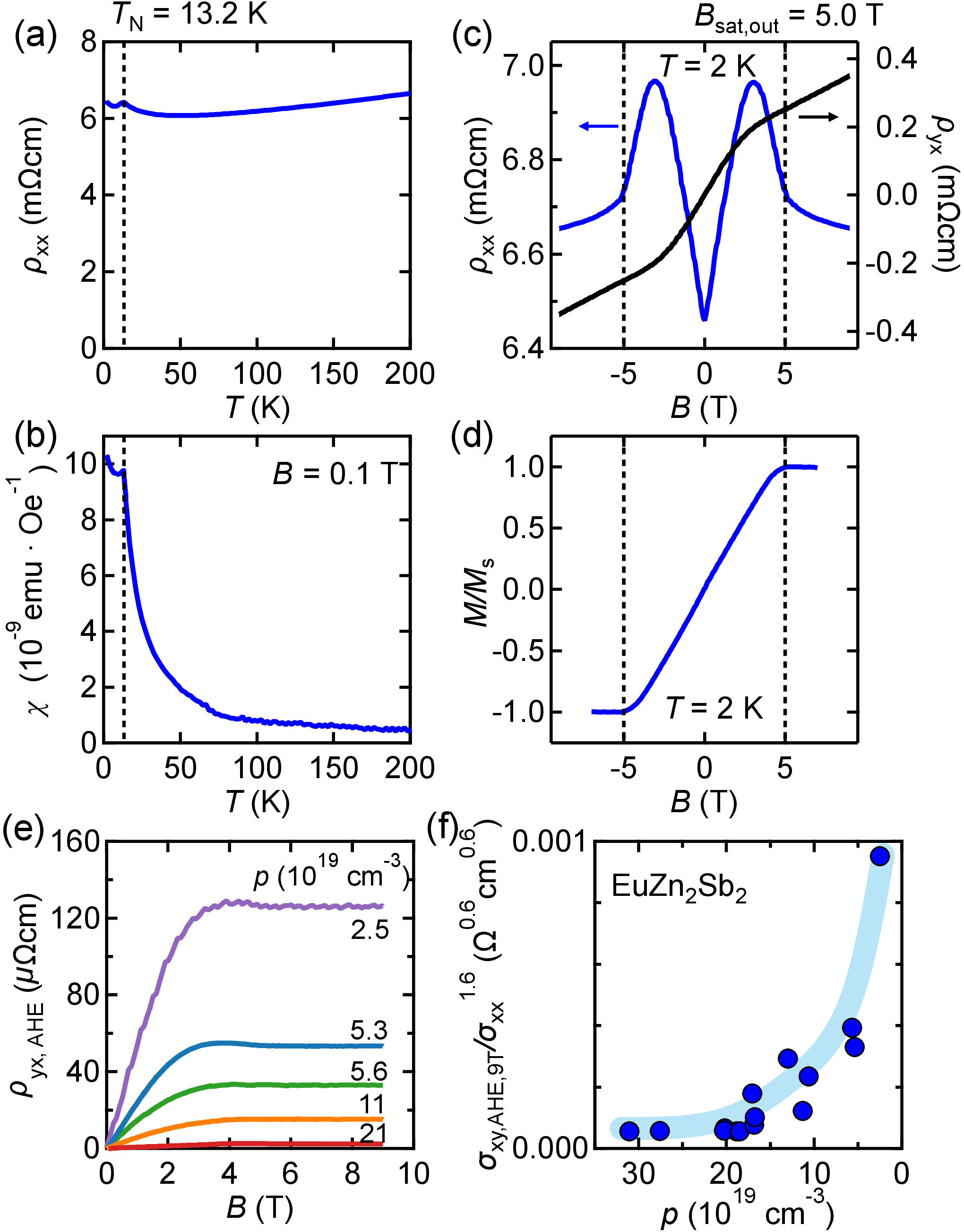}%
\caption{Fundamental magnetotransport of \ce{EuZn2Sb2} thin films.
Temperature dependences of (a) longitudinal resistivity $\rho_\text{xx}$ and (b) magnetic susceptibility $\chi $.
The N\'eel temperature $T_{\text{N}}$ of 13.2 K is indicated by a dashed line.
(c) Magnetic field dependence of $\rho_\text{xx}$ and Hall resistivity $\rho_\text{yx}$, and (d) a magnetization curve, 
taken under the out-of-plane magnetic field at 2 K. 
The out-of-plan saturation field $B_\text{sat,out}$ is determined to be 5.0 T.
(e) Anomalous Hall resistivity $\rho_\text{yx,AHE}$ extracted for \ce{EuZn2Sb2} films with different hole densities is at 2 K.
(f) Hole-density dependence of $\sigma_\text{xy,AHE,9T}/\sigma_\text{xx}^{1.6}$, clearly revealing the large enhancement with decrease in the hole density.}
\label{fig2} 
\end{center}
\end{figure}

To understand the fundamental magnetotransport of \ce{EuZn2Sb2} thin films, we first present their transport and magnetization measured under the out-of-plane field. Temperature dependences of the longitudinal resistivity $\rho_\text{xx}$ in Fig. \ref{fig2}(a) and the magnetic susceptibility $\chi $ in Fig. \ref{fig2}(b) show a kink at 13.2 K, which is in good agreement with the N\'eel temperature $T_\text{N}$ previously reported for \ce{EuZn2Sb2} bulk samples \cite{weber2006low,zhang2008new,singh2024large,sprague2024observation}. Especially above $T_\text{N}$, $\rho_\text{xx}$ shows a metallic behavior. Application of an out-of-plane magnetic field below $T_\text{N}$ leads to a simple magnetization curve where the antiferromagnetically ordered in-plane spin magnetic moments are gradually canted to the out-of-plane field direction. From both the magnetoresistivity curve in Fig. \ref{fig2}(c) and the magnetization curve in Fig. \ref{fig2}(d), the saturation field is estimated at 5.0 T. In the Hall resistivity $\rho_\text{yx}$ shown in Fig. \ref{fig2}(c), the out-of-plane anomalous Hall effect is clearly observed even before subtracting the ordinary Hall component. Figure \ref{fig2}(e) shows the anomalous Hall resistivity $\rho_\text{yx,AHE}$ for \ce{EuZn2Sb2} films with different carrier densities, obtained by subtracting the field-linear term from $\rho_\text{yx}$. $\rho_\text{yx,AHE}$ rapidly increases upon decreasing hole density. \textcolor{black}{To confirm this trend more carefully, in Fig. \ref{fig2}(f) we plot the hole-density dependence of $\sigma_\text{xy,AHE,9T}/\sigma_\text{xx}^{1.6}$ calculated from the converted longitudinal conductivity $\sigma_\text{xx}$ and the anomalous Hall conductivity at 9T $\sigma_\text{xy,AHE,9T}$, following the scaling relationship of the dirty metal region with impurity scattering \cite{onoda2008quantum}.} In contrast to the case of the magnetic Weyl semimetal \ce{EuCd2Sb2}, where $\sigma_\text{xy,AHE,9T}/\sigma_\text{xx}^{1.6}$ is maximized with the Fermi energy tuned to the Weyl points energy range \cite{ohno2022maximizing,su2020magnetic}, $\sigma_\text{xy,AHE,9T}/\sigma_\text{xx}^{1.6}$ continues to increase with decrease in the hole density in \ce{EuZn2Sb2}. This suggests the presence of Berry curvature hotspots in the higher Fermi energy range in \ce{EuZn2Sb2}.

\begin{figure}
\begin{center}
\includegraphics[width=17.5cm]{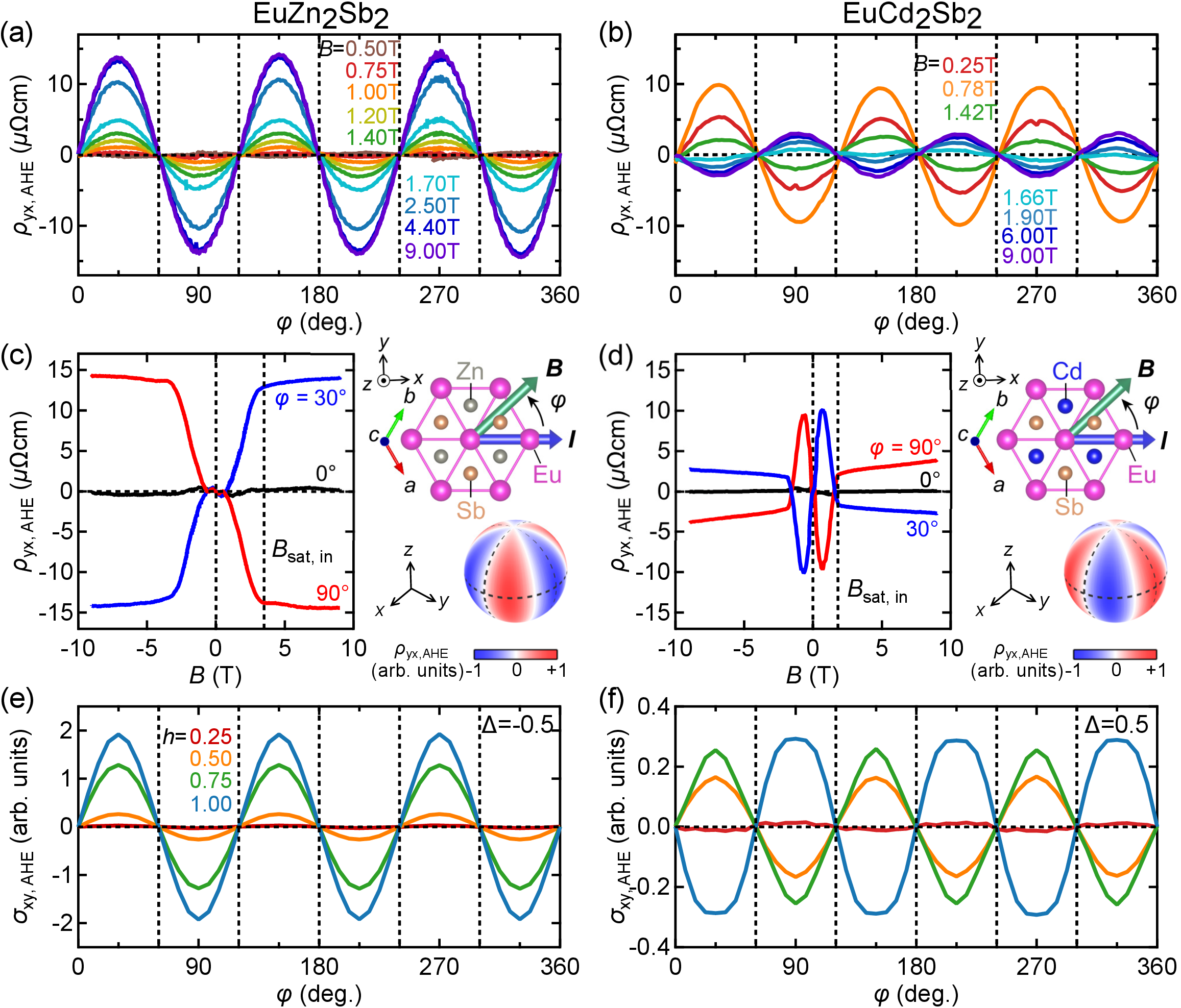}%
\caption{Comparison of the in-plane anomalous Hall resistivity between \ce{EuZn2Sb2} and \ce{EuCd2Sb2}.
In-plane field angle dependence of $\rho_\text{yx,AHE}$ taken for 
(a) \ce{EuZn2Sb2} and (b) \ce{EuCd2Sb2} films at various magnetic fields.
In-plane field scans of $\rho_\text{yx,AHE}$ for the (c) \ce{EuZn2Sb2} and (d) \ce{EuCd2Sb2} films at specific angles of $\varphi= 0\degree$, $30\degree$, and $90\degree$.
The schematic on the right illustrates the measurement configuration with respect to the crystal axes and field angle dependence of $\rho_\text{yx,AHE}$ observed above $B_\text{sat,in}$. 
In-plane field angle dependence of the anomalous Hall conductivity $\sigma_{\text{xy,AHE}}$,
calculated for the band splitting parameters (e) $\Delta = -0.5$ and (f) $\Delta = 0.5$ at different fields $h=B/B_\text{sat,in}$ below $B_\text{sat,in}$.}
\label{fig3}
\end{center}
\end{figure}

Next, we move on to magnetotransport under the in-plane magnetic field. The electric current is applied along the [110] direction of \ce{EuZn2Sb2} and the in-plane magnetic field is rotated with the azimuthal angle $\varphi$ measured from the [110] direction as shown in Fig. \ref{fig3}. We analyzed the raw $\varphi$-scan data to separate different in-plane phenomena as like the two-fold planar Hall effect (PHE) and the three-fold in-plane anomalous Hall effect \cite{kurumaji2023symmetry}. \textcolor{black}{A comprehensive data analysis of PHE is provided in the Supplemental Material \cite{supp}. As detailed therein, we observed a sign reversal in the PHE resistivity ($\rho_\text{yx,PHE}$) and the anisotropic magnetoresistivity (AMR) ($\rho_\text{xx,AMR}$) \cite{thomson1857xix,mcguire1975anisotropic,burkov2017giant,nandy2017chiral,kumar2018planar}.}

Observed iAHE can be explained by the orbital magnetization model as in \ce{EuCd2Sb2} \cite{nakamura2024observation}, where the out-of-plane Berry curvature and orbital magnetization components are induced by the in-plane magnetic field. The octupole model considering the higher order of spin magnetization is not applicable in the present case, since out-of-plane tilting of the Hall vector by this effect occurs only when the measurement plane is not a principal plane \cite{peng2024observation}. In addition, according to the anomalous orbital polarizability model \cite{wang2024orbital}, tensor components responsible for iAHE are zero in $D_{\text{3}d}$ systems. Figure \ref{fig3}(c) presents the in-plane field dependence of $\rho_\text{yx,AHE}$ at $\varphi=30\degree, $. This field dependence seen in $\rho_\text{yx,AHE}$ is completely different from the $\rho_\text{yx,PHE}$ one. It is also noticeable that $\rho_{\text{yx,AHE}}$ keeps increasing even above $B_\text{sat,in}$. This is because the Zeeman-type band splitting continues to increase and the resulting orbital magnetization keeps changing even above the saturation field of the spin magnetization \cite{burkov2011weyl,ito2017anomalous,nakamura2024observation}.

Figure \ref{fig3} shows a comparison of the iAHE between the \ce{EuZn2Sb2} and \ce{EuCd2Sb2} films. \textcolor{black}{The azimuthal angle $\varphi$ is measured from the [110] direction as illustrated in the inset of Fig. \ref{fig3}, where the Zn/Cd atoms are located in the same positions in \ce{EuZn2Sb2}/\ce{EuCd2Sb2}.} Primarily there are two important differences: Firstly, the sign of iAHE especially above $B_{\text{sat,in}}$ is opposite in the two compounds. As confirmed in Figs. \ref{fig3}(a) and \ref{fig3}(b), the 9-T curve is negative, for example, at $\varphi=30\degree$ in \ce{EuCd2Sb2} but positive in \ce{EuZn2Sb2}. This can be confirmed also in the magnetic field scans in Figs. \ref{fig3}(c) and \ref{fig3}(d), suggesting the possibility that the out-of-plane orbital magnetization points in opposite directions between the two compounds. Secondly, the field dependence of the iAHE below $B_{\text{sat,in}}$ is different. In the \ce{EuCd2Sb2} film, $\rho_\text{yx,AHE}$ exhibits a non-monotonic behavior characterized by a pronounced peak at 0.7 T. This peak can be explained by the non-monotonic change of the Berry curvature during the formation and shifting of the Weyl points near the Fermi energy during the magnetization process \cite{takahashi2018anomalous,nakamura2024berry}. On the other hand, such non-monotonic dependence is hardly seen in \ce{EuZn2Sb2}.

To identify the origin of the differences in iAHE, we have performed the calculation of anomalous Hall conductivity using a microscopic model. \textcolor{black}{According to previous studies \cite{soh2018magnetic}, the bands of \ce{EuCd2Sb2} near the Fermi level at $\Gamma$ point consist of \ce{Cd} 5$s$ bands and Sb 5$p$ bands, where the conduction band mainly consists of \ce{Cd} $^2S_{\text{1/2}}$ and \ce{Sb} $^2P_{\text{1/2}}$ orbitals, and the \ce{Sb} $^2P_{\text{3/2}}$ orbitals construct the valence band. As the Fermi level of our \ce{EuZn2Sb2} thin films lays in the valence band, here we consider the four valence bands arising from the \ce{Sb} $^2P_{\text{3/2}}$ orbitals for our model \cite{murakami20042}.} The effective model near the $\Gamma$ point is
\begin{equation}
H_0=d_0(\overrightarrow{k})+\sum_\text{i=1}^5d_i(\overrightarrow{k})\Gamma_i,
\end{equation}
where $d_0(\overrightarrow{k})=\mu +a^0_zk^2_z+a^0_\perp (k^2_z+k^2_y)$, $d_1(\overrightarrow{k} )=\sqrt{3}\left[E_\text{11} k_y k_z+E_\text{21}(k_y^2-k_x^2)\right] $,  $d_2(\overrightarrow{k} )= \sqrt{3}\left[E_\text{11}k_x k_z - 2E_\text{21}K_x k_y\right] $, $d_3(\overrightarrow{k}) = \sqrt{3}\left[-E_\text{12}k_x k_z+2E_\text{22}k_x k_y\right] $, $d_4(\overrightarrow{k}) = \sqrt{3} \left[E_\text{22}(k^2_x - k^2_y)-E_\text{12}k_y k_z\right] $, and $d_5(\overrightarrow{k} )= \Delta + a^z_zk^2_z +a^z_\perp (k_z^2+k^2_y)$ \cite{nakamura2024observation}. The matrices $\Gamma_\text{1}=\sigma^\text{3}\otimes \sigma^\text{2}$, $\Gamma_\text{2}=\sigma^\text{3}\otimes \sigma^\text{1}$, $\Gamma_\text{3}=\sigma^\text{2}\otimes \sigma^\text{0}$, $\Gamma_\text{4}=\sigma^\text{1}\otimes \sigma^\text{0}$, and $\Gamma_\text{5}=\sigma^\text{3}\otimes \sigma^\text{3}$ are time-reversal symmetric $4\times 4$ matrices \cite{murakami20042}, where $\sigma^\text{1,2,3}$ are the Pauli matrices, and $\sigma^0$ is the $2\times 2$ unit matrix. 

\textcolor{black}{The bands hosting the Weyl point pairs near the $\Gamma$ point are dominated by the \ce{Sb} $^2P_\text{3/2}$ orbitals in \ce{EuCd2Sb2} \cite{soh2018magnetic}. The effect of the chemical substitution between \ce{Cd} and \ce{Zn} results in the change of lattice parameter, especially the uniaxial anisotropy $c/a$. The band inversion is modulated by the change of uniaxial anisotropy, which is effectively expressed by the parameter $\Delta$.} To understand the effect of $\Delta$, we compare the anomalous Hall conductivity $\sigma_\text{xy,AHE}$ by simply changing the band splitting parameter $\Delta$; the other parameters are fixed to $\mu =-1.5$, $a_z^0 = a^0_\perp =-2.0$, $a^z_z = -0.5$, $a^z_\perp =0.25$, $E_{11}=-E_{21}=-0.5$, and $E_{12}=E_{22}=0$. The calculated $\varphi$ dependence for $\Delta = -0.5$ and $0.5$ are shown in Fig. \ref{fig3}(e)  and (f) corresponds to an inverted band structure with Weyl points. $\sigma_\text{xy,AHE}$ calculated for $\Delta = -0.5$ is positive at $\varphi = 30\degree$ and monotonically increases with increase in the in-plane magnetic field. On the other hand, $\sigma_\text{xy,AHE}$ calculated for $\Delta = 0.5$ non-monotonically changes with sign inversion and exhibits a negative value at the saturation field ($h=B/B_\text{sat,in}=1$). These features found for $\Delta = -0.5$ and $0.5$ are quite similar to the observations in \ce{EuZn2Sb2} and \ce{EuCd2Sb2}, respectively.

\begin{figure}
\begin{center}
\includegraphics[width=11.0cm]{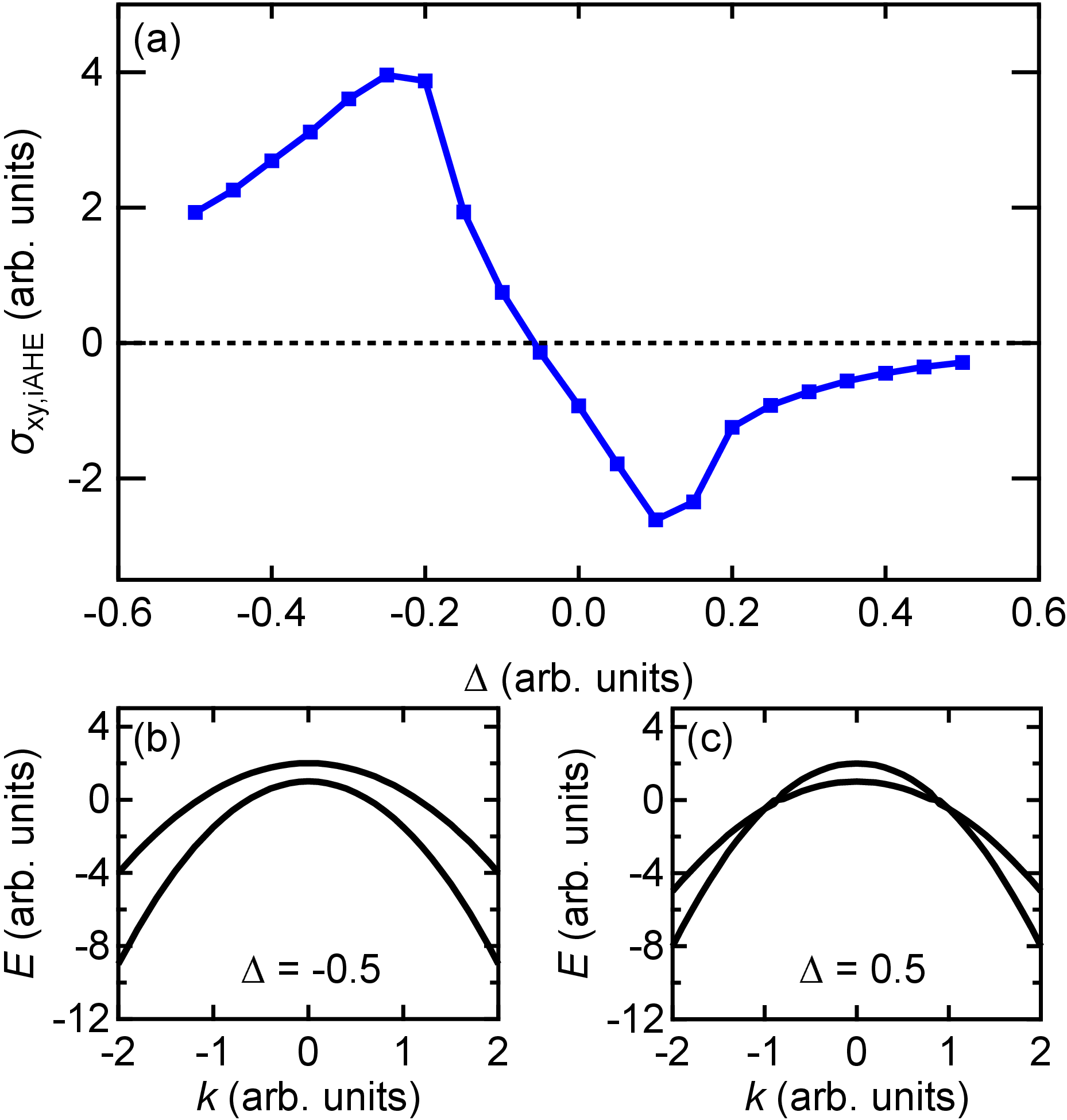}%
\caption{\textcolor{black}{Relation between in-plane anomalous Hall conductivity and the parameter $\Delta$ representing the band modulation through the uniaxial anisotropy.
(a) $\Delta$ dependence of in-plane anomalous Hall conductivity at $\varphi=30\degree$
Band dispersions with (b) $\Delta=-0.5$ and (c) $\Delta=0.5$.
}}
\label{fig4}
\end{center}
\end{figure}

\textcolor{black}{To see the effect of $\Delta$ more clearly, we present the $\Delta$ dependence of anomalous Hall conductivity $\sigma_\text{xy,AHE}$. The band dispersion with $\Delta = -0.5$ and $\Delta = 0.5$ are also shown in Figs. \ref{fig4}(b) and \ref{fig4}(c), respectively. There is a band crossing between the Sb 5$p$ bands in the case of $\Delta = 0.5$, while there is not for $\Delta = -0.5$. The anomalous Hall conductivity changes its sign around the region where the band inversion is lifted. By changing the band inversion parameter $\Delta$ related to uniaxial anisotropy, the drastic change in the iAHE response as confirmed between \ce{EuZn2Sb2} and \ce{EuCd2Sb2} can be simply demonstrated.}

In summary, we have studied the in-plane anomalous Hall effect in single-crystalline films of \ce{EuZn2Sb2}, which is a good reference material with a weaker SOC than in \ce{EuCd2Sb2}. The in-plane anomalous Hall effect clearly appears with three-fold azimuthal angle dependence in addition to the conventional two-fold planar Hall effect. In comparison to \ce{EuCd2Sb2}, \ce{EuZn2Sb2} exhibits the in-plane anomalous Hall resistivity with opposite sign and rather a monotonic field dependence. These differences are qualitatively reproduced in our model calculations by simply changing the band inversion parameter caused by uniaxial anisotropy. Our findings lay the groundwork for systematically controlling in-plane anomalous Hall effect and orbital magnetization through elaborate band engineering.

\begin{acknowledgments}
This work was supported by JST FOREST Program Grant No. JPMJFR202N, Japan, by Grant-in-Aids for Scientific Research JP22K18967, JP22K20353, JP23K13666, JP23K03275, JP24H01614, and JP24H01654 from MEXT, Japan, by Murata Science and Education Foundation, Japan, by Iketani Science and Technology Foundation, Japan, and by STAR Award funded by the Tokyo Tech Fund, Japan.
\end{acknowledgments}


\begin{thebibliography}{44}%
\makeatletter
\providecommand \@ifxundefined [1]{%
 \@ifx{#1\undefined}
}%
\providecommand \@ifnum [1]{%
 \ifnum #1\expandafter \@firstoftwo
 \else \expandafter \@secondoftwo
 \fi
}%
\providecommand \@ifx [1]{%
 \ifx #1\expandafter \@firstoftwo
 \else \expandafter \@secondoftwo
 \fi
}%
\providecommand \natexlab [1]{#1}%
\providecommand \enquote  [1]{``#1''}%
\providecommand \bibnamefont  [1]{#1}%
\providecommand \bibfnamefont [1]{#1}%
\providecommand \citenamefont [1]{#1}%
\providecommand \href@noop [0]{\@secondoftwo}%
\providecommand \href [0]{\begingroup \@sanitize@url \@href}%
\providecommand \@href[1]{\@@startlink{#1}\@@href}%
\providecommand \@@href[1]{\endgroup#1\@@endlink}%
\providecommand \@sanitize@url [0]{\catcode `\\12\catcode `\$12\catcode `\&12\catcode `\#12\catcode `\^12\catcode `\_12\catcode `\%12\relax}%
\providecommand \@@startlink[1]{}%
\providecommand \@@endlink[0]{}%
\providecommand \url  [0]{\begingroup\@sanitize@url \@url }%
\providecommand \@url [1]{\endgroup\@href {#1}{\urlprefix }}%
\providecommand \urlprefix  [0]{URL }%
\providecommand \Eprint [0]{\href }%
\providecommand \doibase [0]{https://doi.org/}%
\providecommand \selectlanguage [0]{\@gobble}%
\providecommand \bibinfo  [0]{\@secondoftwo}%
\providecommand \bibfield  [0]{\@secondoftwo}%
\providecommand \translation [1]{[#1]}%
\providecommand \BibitemOpen [0]{}%
\providecommand \bibitemStop [0]{}%
\providecommand \bibitemNoStop [0]{.\EOS\space}%
\providecommand \EOS [0]{\spacefactor3000\relax}%
\providecommand \BibitemShut  [1]{\csname bibitem#1\endcsname}%
\let\auto@bib@innerbib\@empty
\bibitem [{\citenamefont {Nagaosa}\ \emph {et~al.}(2010)\citenamefont {Nagaosa}, \citenamefont {Sinova}, \citenamefont {Onoda}, \citenamefont {MacDonald},\ and\ \citenamefont {Ong}}]{nagaosa2010anomalous}%
  \BibitemOpen
  \bibfield  {author} {\bibinfo {author} {\bibfnamefont {N.}~\bibnamefont {Nagaosa}}, \bibinfo {author} {\bibfnamefont {J.}~\bibnamefont {Sinova}}, \bibinfo {author} {\bibfnamefont {S.}~\bibnamefont {Onoda}}, \bibinfo {author} {\bibfnamefont {A.~H.}\ \bibnamefont {MacDonald}},\ and\ \bibinfo {author} {\bibfnamefont {N.~P.}\ \bibnamefont {Ong}},\ }\href@noop {} {\bibfield  {journal} {\bibinfo  {journal} {Reviews of Modern Physics}\ }\textbf {\bibinfo {volume} {82}},\ \bibinfo {pages} {1539} (\bibinfo {year} {2010})}\BibitemShut {NoStop}%
\bibitem [{\citenamefont {Fang}\ \emph {et~al.}(2003)\citenamefont {Fang}, \citenamefont {Nagaosa}, \citenamefont {Takahashi}, \citenamefont {Asamitsu}, \citenamefont {Mathieu}, \citenamefont {Ogasawara}, \citenamefont {Yamada}, \citenamefont {Kawasaki}, \citenamefont {Tokura},\ and\ \citenamefont {Terakura}}]{fang2003anomalous}%
  \BibitemOpen
  \bibfield  {author} {\bibinfo {author} {\bibfnamefont {Z.}~\bibnamefont {Fang}}, \bibinfo {author} {\bibfnamefont {N.}~\bibnamefont {Nagaosa}}, \bibinfo {author} {\bibfnamefont {K.~S.}\ \bibnamefont {Takahashi}}, \bibinfo {author} {\bibfnamefont {A.}~\bibnamefont {Asamitsu}}, \bibinfo {author} {\bibfnamefont {R.}~\bibnamefont {Mathieu}}, \bibinfo {author} {\bibfnamefont {T.}~\bibnamefont {Ogasawara}}, \bibinfo {author} {\bibfnamefont {H.}~\bibnamefont {Yamada}}, \bibinfo {author} {\bibfnamefont {M.}~\bibnamefont {Kawasaki}}, \bibinfo {author} {\bibfnamefont {Y.}~\bibnamefont {Tokura}},\ and\ \bibinfo {author} {\bibfnamefont {K.}~\bibnamefont {Terakura}},\ }\href@noop {} {\bibfield  {journal} {\bibinfo  {journal} {Science}\ }\textbf {\bibinfo {volume} {302}},\ \bibinfo {pages} {92} (\bibinfo {year} {2003})}\BibitemShut {NoStop}%
\bibitem [{\citenamefont {Yao}\ \emph {et~al.}(2004)\citenamefont {Yao}, \citenamefont {Kleinman}, \citenamefont {MacDonald}, \citenamefont {Sinova}, \citenamefont {Jungwirth}, \citenamefont {Wang}, \citenamefont {Wang},\ and\ \citenamefont {Niu}}]{yao2004first}%
  \BibitemOpen
  \bibfield  {author} {\bibinfo {author} {\bibfnamefont {Y.}~\bibnamefont {Yao}}, \bibinfo {author} {\bibfnamefont {L.}~\bibnamefont {Kleinman}}, \bibinfo {author} {\bibfnamefont {A.}~\bibnamefont {MacDonald}}, \bibinfo {author} {\bibfnamefont {J.}~\bibnamefont {Sinova}}, \bibinfo {author} {\bibfnamefont {T.}~\bibnamefont {Jungwirth}}, \bibinfo {author} {\bibfnamefont {D.-s.}\ \bibnamefont {Wang}}, \bibinfo {author} {\bibfnamefont {E.}~\bibnamefont {Wang}},\ and\ \bibinfo {author} {\bibfnamefont {Q.}~\bibnamefont {Niu}},\ }\href@noop {} {\bibfield  {journal} {\bibinfo  {journal} {Physical Review Letters}\ }\textbf {\bibinfo {volume} {92}},\ \bibinfo {pages} {037204} (\bibinfo {year} {2004})}\BibitemShut {NoStop}%
\bibitem [{\citenamefont {Nakamura}\ \emph {et~al.}(2024{\natexlab{a}})\citenamefont {Nakamura}, \citenamefont {Nishihaya}, \citenamefont {Ishizuka}, \citenamefont {Kriener}, \citenamefont {Watanabe},\ and\ \citenamefont {Uchida}}]{nakamura2024observation}%
  \BibitemOpen
  \bibfield  {author} {\bibinfo {author} {\bibfnamefont {A.}~\bibnamefont {Nakamura}}, \bibinfo {author} {\bibfnamefont {S.}~\bibnamefont {Nishihaya}}, \bibinfo {author} {\bibfnamefont {H.}~\bibnamefont {Ishizuka}}, \bibinfo {author} {\bibfnamefont {M.}~\bibnamefont {Kriener}}, \bibinfo {author} {\bibfnamefont {Y.}~\bibnamefont {Watanabe}},\ and\ \bibinfo {author} {\bibfnamefont {M.}~\bibnamefont {Uchida}},\ }\href@noop {} {\bibfield  {journal} {\bibinfo  {journal} {Physical Review Letters}\ }\textbf {\bibinfo {volume} {133}},\ \bibinfo {pages} {236602} (\bibinfo {year} {2024}{\natexlab{a}})}\BibitemShut {NoStop}%
\bibitem [{\citenamefont {Peng}\ \emph {et~al.}(2024)\citenamefont {Peng}, \citenamefont {Liu}, \citenamefont {Pan}, \citenamefont {Wang}, \citenamefont {Chen}, \citenamefont {Zhang}, \citenamefont {Yu}, \citenamefont {Shen}, \citenamefont {Yang}, \citenamefont {Niu} \emph {et~al.}}]{peng2024observation}%
  \BibitemOpen
  \bibfield  {author} {\bibinfo {author} {\bibfnamefont {W.}~\bibnamefont {Peng}}, \bibinfo {author} {\bibfnamefont {Z.}~\bibnamefont {Liu}}, \bibinfo {author} {\bibfnamefont {H.}~\bibnamefont {Pan}}, \bibinfo {author} {\bibfnamefont {P.}~\bibnamefont {Wang}}, \bibinfo {author} {\bibfnamefont {Y.}~\bibnamefont {Chen}}, \bibinfo {author} {\bibfnamefont {J.}~\bibnamefont {Zhang}}, \bibinfo {author} {\bibfnamefont {X.}~\bibnamefont {Yu}}, \bibinfo {author} {\bibfnamefont {J.}~\bibnamefont {Shen}}, \bibinfo {author} {\bibfnamefont {M.}~\bibnamefont {Yang}}, \bibinfo {author} {\bibfnamefont {Q.}~\bibnamefont {Niu}}, \emph {et~al.},\ }\href@noop {} {\bibfield  {journal} {\bibinfo  {journal} {arXiv:2402.15741}\ } (\bibinfo {year} {2024})}\BibitemShut {NoStop}%
\bibitem [{\citenamefont {Liu}\ \emph {et~al.}(2013)\citenamefont {Liu}, \citenamefont {Hsu},\ and\ \citenamefont {Liu}}]{liu2013plane}%
  \BibitemOpen
  \bibfield  {author} {\bibinfo {author} {\bibfnamefont {X.}~\bibnamefont {Liu}}, \bibinfo {author} {\bibfnamefont {H.-C.}\ \bibnamefont {Hsu}},\ and\ \bibinfo {author} {\bibfnamefont {C.-X.}\ \bibnamefont {Liu}},\ }\href@noop {} {\bibfield  {journal} {\bibinfo  {journal} {Physical Review Letters}\ }\textbf {\bibinfo {volume} {111}},\ \bibinfo {pages} {086802} (\bibinfo {year} {2013})}\BibitemShut {NoStop}%
\bibitem [{\citenamefont {Ren}\ \emph {et~al.}(2016)\citenamefont {Ren}, \citenamefont {Zeng}, \citenamefont {Deng}, \citenamefont {Yang}, \citenamefont {Pan},\ and\ \citenamefont {Qiao}}]{ren2016quantum}%
  \BibitemOpen
  \bibfield  {author} {\bibinfo {author} {\bibfnamefont {Y.}~\bibnamefont {Ren}}, \bibinfo {author} {\bibfnamefont {J.}~\bibnamefont {Zeng}}, \bibinfo {author} {\bibfnamefont {X.}~\bibnamefont {Deng}}, \bibinfo {author} {\bibfnamefont {F.}~\bibnamefont {Yang}}, \bibinfo {author} {\bibfnamefont {H.}~\bibnamefont {Pan}},\ and\ \bibinfo {author} {\bibfnamefont {Z.}~\bibnamefont {Qiao}},\ }\href@noop {} {\bibfield  {journal} {\bibinfo  {journal} {Physical Review B}\ }\textbf {\bibinfo {volume} {94}},\ \bibinfo {pages} {085411} (\bibinfo {year} {2016})}\BibitemShut {NoStop}%
\bibitem [{\citenamefont {Zhang}\ \emph {et~al.}(2019)\citenamefont {Zhang}, \citenamefont {Liu},\ and\ \citenamefont {Wang}}]{zhang2019plane}%
  \BibitemOpen
  \bibfield  {author} {\bibinfo {author} {\bibfnamefont {J.}~\bibnamefont {Zhang}}, \bibinfo {author} {\bibfnamefont {Z.}~\bibnamefont {Liu}},\ and\ \bibinfo {author} {\bibfnamefont {J.}~\bibnamefont {Wang}},\ }\href@noop {} {\bibfield  {journal} {\bibinfo  {journal} {Physical Review B}\ }\textbf {\bibinfo {volume} {100}},\ \bibinfo {pages} {165117} (\bibinfo {year} {2019})}\BibitemShut {NoStop}%
\bibitem [{\citenamefont {Ren}\ \emph {et~al.}(2020)\citenamefont {Ren}, \citenamefont {Qiao},\ and\ \citenamefont {Niu}}]{ren2020engineering}%
  \BibitemOpen
  \bibfield  {author} {\bibinfo {author} {\bibfnamefont {Y.}~\bibnamefont {Ren}}, \bibinfo {author} {\bibfnamefont {Z.}~\bibnamefont {Qiao}},\ and\ \bibinfo {author} {\bibfnamefont {Q.}~\bibnamefont {Niu}},\ }\href@noop {} {\bibfield  {journal} {\bibinfo  {journal} {Physical Review Letters}\ }\textbf {\bibinfo {volume} {124}},\ \bibinfo {pages} {166804} (\bibinfo {year} {2020})}\BibitemShut {NoStop}%
\bibitem [{\citenamefont {Sun}\ \emph {et~al.}(2022)\citenamefont {Sun}, \citenamefont {Weng},\ and\ \citenamefont {Dai}}]{sun2022possible}%
  \BibitemOpen
  \bibfield  {author} {\bibinfo {author} {\bibfnamefont {S.}~\bibnamefont {Sun}}, \bibinfo {author} {\bibfnamefont {H.}~\bibnamefont {Weng}},\ and\ \bibinfo {author} {\bibfnamefont {X.}~\bibnamefont {Dai}},\ }\href@noop {} {\bibfield  {journal} {\bibinfo  {journal} {Physical Review B}\ }\textbf {\bibinfo {volume} {106}},\ \bibinfo {pages} {L241105} (\bibinfo {year} {2022})}\BibitemShut {NoStop}%
\bibitem [{\citenamefont {Li}\ \emph {et~al.}(2022)\citenamefont {Li}, \citenamefont {Han},\ and\ \citenamefont {Qiao}}]{li2022chern}%
  \BibitemOpen
  \bibfield  {author} {\bibinfo {author} {\bibfnamefont {Z.}~\bibnamefont {Li}}, \bibinfo {author} {\bibfnamefont {Y.}~\bibnamefont {Han}},\ and\ \bibinfo {author} {\bibfnamefont {Z.}~\bibnamefont {Qiao}},\ }\href@noop {} {\bibfield  {journal} {\bibinfo  {journal} {Physical Review Letters}\ }\textbf {\bibinfo {volume} {129}},\ \bibinfo {pages} {036801} (\bibinfo {year} {2022})}\BibitemShut {NoStop}%
\bibitem [{\citenamefont {Li}\ \emph {et~al.}(2023)\citenamefont {Li}, \citenamefont {Cao}, \citenamefont {Cui}, \citenamefont {Yu},\ and\ \citenamefont {Yao}}]{li2023planar}%
  \BibitemOpen
  \bibfield  {author} {\bibinfo {author} {\bibfnamefont {L.}~\bibnamefont {Li}}, \bibinfo {author} {\bibfnamefont {J.}~\bibnamefont {Cao}}, \bibinfo {author} {\bibfnamefont {C.}~\bibnamefont {Cui}}, \bibinfo {author} {\bibfnamefont {Z.-M.}\ \bibnamefont {Yu}},\ and\ \bibinfo {author} {\bibfnamefont {Y.}~\bibnamefont {Yao}},\ }\href@noop {} {\bibfield  {journal} {\bibinfo  {journal} {Physical Review B}\ }\textbf {\bibinfo {volume} {108}},\ \bibinfo {pages} {085120} (\bibinfo {year} {2023})}\BibitemShut {NoStop}%
\bibitem [{\citenamefont {Cao}\ \emph {et~al.}(2023)\citenamefont {Cao}, \citenamefont {Jiang}, \citenamefont {Li}, \citenamefont {Tu}, \citenamefont {Zhou}, \citenamefont {Zhou},\ and\ \citenamefont {Yao}}]{cao2023plane}%
  \BibitemOpen
  \bibfield  {author} {\bibinfo {author} {\bibfnamefont {J.}~\bibnamefont {Cao}}, \bibinfo {author} {\bibfnamefont {W.}~\bibnamefont {Jiang}}, \bibinfo {author} {\bibfnamefont {X.-P.}\ \bibnamefont {Li}}, \bibinfo {author} {\bibfnamefont {D.}~\bibnamefont {Tu}}, \bibinfo {author} {\bibfnamefont {J.}~\bibnamefont {Zhou}}, \bibinfo {author} {\bibfnamefont {J.}~\bibnamefont {Zhou}},\ and\ \bibinfo {author} {\bibfnamefont {Y.}~\bibnamefont {Yao}},\ }\href@noop {} {\bibfield  {journal} {\bibinfo  {journal} {Physical Review Letters}\ }\textbf {\bibinfo {volume} {130}},\ \bibinfo {pages} {166702} (\bibinfo {year} {2023})}\BibitemShut {NoStop}%
\bibitem [{\citenamefont {Liang}\ \emph {et~al.}(2018)\citenamefont {Liang}, \citenamefont {Lin}, \citenamefont {Gibson}, \citenamefont {Kushwaha}, \citenamefont {Liu}, \citenamefont {Wang}, \citenamefont {Xiong}, \citenamefont {Sobota}, \citenamefont {Hashimoto}, \citenamefont {Kirchmann} \emph {et~al.}}]{liang2018anomalous}%
  \BibitemOpen
  \bibfield  {author} {\bibinfo {author} {\bibfnamefont {T.}~\bibnamefont {Liang}}, \bibinfo {author} {\bibfnamefont {J.}~\bibnamefont {Lin}}, \bibinfo {author} {\bibfnamefont {Q.}~\bibnamefont {Gibson}}, \bibinfo {author} {\bibfnamefont {S.}~\bibnamefont {Kushwaha}}, \bibinfo {author} {\bibfnamefont {M.}~\bibnamefont {Liu}}, \bibinfo {author} {\bibfnamefont {W.}~\bibnamefont {Wang}}, \bibinfo {author} {\bibfnamefont {H.}~\bibnamefont {Xiong}}, \bibinfo {author} {\bibfnamefont {J.~A.}\ \bibnamefont {Sobota}}, \bibinfo {author} {\bibfnamefont {M.}~\bibnamefont {Hashimoto}}, \bibinfo {author} {\bibfnamefont {P.~S.}\ \bibnamefont {Kirchmann}}, \emph {et~al.},\ }\href@noop {} {\bibfield  {journal} {\bibinfo  {journal} {Nature Physics}\ }\textbf {\bibinfo {volume} {14}},\ \bibinfo {pages} {451} (\bibinfo {year} {2018})}\BibitemShut {NoStop}%
\bibitem [{\citenamefont {Zhou}\ \emph {et~al.}(2022)\citenamefont {Zhou}, \citenamefont {Zhang}, \citenamefont {Lin}, \citenamefont {Cao}, \citenamefont {Zhou}, \citenamefont {Jiang}, \citenamefont {Du}, \citenamefont {Tang}, \citenamefont {Shi}, \citenamefont {Jiang} \emph {et~al.}}]{zhou2022heterodimensional}%
  \BibitemOpen
  \bibfield  {author} {\bibinfo {author} {\bibfnamefont {J.}~\bibnamefont {Zhou}}, \bibinfo {author} {\bibfnamefont {W.}~\bibnamefont {Zhang}}, \bibinfo {author} {\bibfnamefont {Y.-C.}\ \bibnamefont {Lin}}, \bibinfo {author} {\bibfnamefont {J.}~\bibnamefont {Cao}}, \bibinfo {author} {\bibfnamefont {Y.}~\bibnamefont {Zhou}}, \bibinfo {author} {\bibfnamefont {W.}~\bibnamefont {Jiang}}, \bibinfo {author} {\bibfnamefont {H.}~\bibnamefont {Du}}, \bibinfo {author} {\bibfnamefont {B.}~\bibnamefont {Tang}}, \bibinfo {author} {\bibfnamefont {J.}~\bibnamefont {Shi}}, \bibinfo {author} {\bibfnamefont {B.}~\bibnamefont {Jiang}}, \emph {et~al.},\ }\href@noop {} {\bibfield  {journal} {\bibinfo  {journal} {Nature}\ }\textbf {\bibinfo {volume} {609}},\ \bibinfo {pages} {46} (\bibinfo {year} {2022})}\BibitemShut {NoStop}%
\bibitem [{\citenamefont {Wang}\ \emph {et~al.}(2024{\natexlab{a}})\citenamefont {Wang}, \citenamefont {Zhu}, \citenamefont {Chen}, \citenamefont {Wang}, \citenamefont {Liu}, \citenamefont {Huang}, \citenamefont {Jiang}, \citenamefont {Zhao}, \citenamefont {Shi}, \citenamefont {Tian} \emph {et~al.}}]{wang2024observation}%
  \BibitemOpen
  \bibfield  {author} {\bibinfo {author} {\bibfnamefont {L.}~\bibnamefont {Wang}}, \bibinfo {author} {\bibfnamefont {J.}~\bibnamefont {Zhu}}, \bibinfo {author} {\bibfnamefont {H.}~\bibnamefont {Chen}}, \bibinfo {author} {\bibfnamefont {H.}~\bibnamefont {Wang}}, \bibinfo {author} {\bibfnamefont {J.}~\bibnamefont {Liu}}, \bibinfo {author} {\bibfnamefont {Y.-X.}\ \bibnamefont {Huang}}, \bibinfo {author} {\bibfnamefont {B.}~\bibnamefont {Jiang}}, \bibinfo {author} {\bibfnamefont {J.}~\bibnamefont {Zhao}}, \bibinfo {author} {\bibfnamefont {H.}~\bibnamefont {Shi}}, \bibinfo {author} {\bibfnamefont {G.}~\bibnamefont {Tian}}, \emph {et~al.},\ }\href@noop {} {\bibfield  {journal} {\bibinfo  {journal} {Physical Review Letters}\ }\textbf {\bibinfo {volume} {132}},\ \bibinfo {pages} {106601} (\bibinfo {year} {2024}{\natexlab{a}})}\BibitemShut {NoStop}%
\bibitem [{\citenamefont {Zyuzin}(2020)}]{zyuzin2020plane}%
  \BibitemOpen
  \bibfield  {author} {\bibinfo {author} {\bibfnamefont {V.~A.}\ \bibnamefont {Zyuzin}},\ }\href@noop {} {\bibfield  {journal} {\bibinfo  {journal} {Physical Review B}\ }\textbf {\bibinfo {volume} {102}},\ \bibinfo {pages} {241105} (\bibinfo {year} {2020})}\BibitemShut {NoStop}%
\bibitem [{\citenamefont {Wang}\ \emph {et~al.}(2024{\natexlab{b}})\citenamefont {Wang}, \citenamefont {Huang}, \citenamefont {Liu}, \citenamefont {Feng}, \citenamefont {Zhu}, \citenamefont {Wu}, \citenamefont {Xiao},\ and\ \citenamefont {Yang}}]{wang2024orbital}%
  \BibitemOpen
  \bibfield  {author} {\bibinfo {author} {\bibfnamefont {H.}~\bibnamefont {Wang}}, \bibinfo {author} {\bibfnamefont {Y.-X.}\ \bibnamefont {Huang}}, \bibinfo {author} {\bibfnamefont {H.}~\bibnamefont {Liu}}, \bibinfo {author} {\bibfnamefont {X.}~\bibnamefont {Feng}}, \bibinfo {author} {\bibfnamefont {J.}~\bibnamefont {Zhu}}, \bibinfo {author} {\bibfnamefont {W.}~\bibnamefont {Wu}}, \bibinfo {author} {\bibfnamefont {C.}~\bibnamefont {Xiao}},\ and\ \bibinfo {author} {\bibfnamefont {S.~A.}\ \bibnamefont {Yang}},\ }\href@noop {} {\bibfield  {journal} {\bibinfo  {journal} {Physical Review Letters}\ }\textbf {\bibinfo {volume} {132}},\ \bibinfo {pages} {056301} (\bibinfo {year} {2024}{\natexlab{b}})}\BibitemShut {NoStop}%
\bibitem [{\citenamefont {Battilomo}\ \emph {et~al.}(2021)\citenamefont {Battilomo}, \citenamefont {Scopigno},\ and\ \citenamefont {Ortix}}]{battilomo2021anomalous}%
  \BibitemOpen
  \bibfield  {author} {\bibinfo {author} {\bibfnamefont {R.}~\bibnamefont {Battilomo}}, \bibinfo {author} {\bibfnamefont {N.}~\bibnamefont {Scopigno}},\ and\ \bibinfo {author} {\bibfnamefont {C.}~\bibnamefont {Ortix}},\ }\href@noop {} {\bibfield  {journal} {\bibinfo  {journal} {Physical Review Research}\ }\textbf {\bibinfo {volume} {3}},\ \bibinfo {pages} {L012006} (\bibinfo {year} {2021})}\BibitemShut {NoStop}%
\bibitem [{\citenamefont {Cullen}\ \emph {et~al.}(2021)\citenamefont {Cullen}, \citenamefont {Bhalla}, \citenamefont {Marcellina}, \citenamefont {Hamilton},\ and\ \citenamefont {Culcer}}]{cullen2021generating}%
  \BibitemOpen
  \bibfield  {author} {\bibinfo {author} {\bibfnamefont {J.~H.}\ \bibnamefont {Cullen}}, \bibinfo {author} {\bibfnamefont {P.}~\bibnamefont {Bhalla}}, \bibinfo {author} {\bibfnamefont {E.}~\bibnamefont {Marcellina}}, \bibinfo {author} {\bibfnamefont {A.~R.}\ \bibnamefont {Hamilton}},\ and\ \bibinfo {author} {\bibfnamefont {D.}~\bibnamefont {Culcer}},\ }\href@noop {} {\bibfield  {journal} {\bibinfo  {journal} {Physical Review Letters}\ }\textbf {\bibinfo {volume} {126}},\ \bibinfo {pages} {256601} (\bibinfo {year} {2021})}\BibitemShut {NoStop}%
\bibitem [{\citenamefont {Roman}\ \emph {et~al.}(2009)\citenamefont {Roman}, \citenamefont {Mokrousov},\ and\ \citenamefont {Souza}}]{roman2009orientation}%
  \BibitemOpen
  \bibfield  {author} {\bibinfo {author} {\bibfnamefont {E.}~\bibnamefont {Roman}}, \bibinfo {author} {\bibfnamefont {Y.}~\bibnamefont {Mokrousov}},\ and\ \bibinfo {author} {\bibfnamefont {I.}~\bibnamefont {Souza}},\ }\href@noop {} {\bibfield  {journal} {\bibinfo  {journal} {Physical Review Letters}\ }\textbf {\bibinfo {volume} {103}},\ \bibinfo {pages} {097203} (\bibinfo {year} {2009})}\BibitemShut {NoStop}%
\bibitem [{\citenamefont {Li}\ \emph {et~al.}(2024)\citenamefont {Li}, \citenamefont {Wang}, \citenamefont {Li},\ and\ \citenamefont {Zhou}}]{li2023switchable}%
  \BibitemOpen
  \bibfield  {author} {\bibinfo {author} {\bibfnamefont {D.}~\bibnamefont {Li}}, \bibinfo {author} {\bibfnamefont {M.}~\bibnamefont {Wang}}, \bibinfo {author} {\bibfnamefont {D.}~\bibnamefont {Li}},\ and\ \bibinfo {author} {\bibfnamefont {J.}~\bibnamefont {Zhou}},\ }\href@noop {} {\bibfield  {journal} {\bibinfo  {journal} {Physical Review B}\ }\textbf {\bibinfo {volume} {109}},\ \bibinfo {pages} {155153} (\bibinfo {year} {2024})}\BibitemShut {NoStop}%
\bibitem [{\citenamefont {Wang}\ \emph {et~al.}(2019)\citenamefont {Wang}, \citenamefont {Jo}, \citenamefont {Kuthanazhi}, \citenamefont {Wu}, \citenamefont {McQueeney}, \citenamefont {Kaminski},\ and\ \citenamefont {Canfield}}]{wang2019single}%
  \BibitemOpen
  \bibfield  {author} {\bibinfo {author} {\bibfnamefont {L.-L.}\ \bibnamefont {Wang}}, \bibinfo {author} {\bibfnamefont {N.~H.}\ \bibnamefont {Jo}}, \bibinfo {author} {\bibfnamefont {B.}~\bibnamefont {Kuthanazhi}}, \bibinfo {author} {\bibfnamefont {Y.}~\bibnamefont {Wu}}, \bibinfo {author} {\bibfnamefont {R.~J.}\ \bibnamefont {McQueeney}}, \bibinfo {author} {\bibfnamefont {A.}~\bibnamefont {Kaminski}},\ and\ \bibinfo {author} {\bibfnamefont {P.~C.}\ \bibnamefont {Canfield}},\ }\href@noop {} {\bibfield  {journal} {\bibinfo  {journal} {Physical Review B}\ }\textbf {\bibinfo {volume} {99}},\ \bibinfo {pages} {245147} (\bibinfo {year} {2019})}\BibitemShut {NoStop}%
\bibitem [{\citenamefont {Su}\ \emph {et~al.}(2020)\citenamefont {Su}, \citenamefont {Gong}, \citenamefont {Shi}, \citenamefont {Yang}, \citenamefont {Wang}, \citenamefont {Xia}, \citenamefont {Yu}, \citenamefont {Guo}, \citenamefont {Wang}, \citenamefont {Ding} \emph {et~al.}}]{su2020magnetic}%
  \BibitemOpen
  \bibfield  {author} {\bibinfo {author} {\bibfnamefont {H.}~\bibnamefont {Su}}, \bibinfo {author} {\bibfnamefont {B.}~\bibnamefont {Gong}}, \bibinfo {author} {\bibfnamefont {W.}~\bibnamefont {Shi}}, \bibinfo {author} {\bibfnamefont {H.}~\bibnamefont {Yang}}, \bibinfo {author} {\bibfnamefont {H.}~\bibnamefont {Wang}}, \bibinfo {author} {\bibfnamefont {W.}~\bibnamefont {Xia}}, \bibinfo {author} {\bibfnamefont {Z.}~\bibnamefont {Yu}}, \bibinfo {author} {\bibfnamefont {P.-J.}\ \bibnamefont {Guo}}, \bibinfo {author} {\bibfnamefont {J.}~\bibnamefont {Wang}}, \bibinfo {author} {\bibfnamefont {L.}~\bibnamefont {Ding}}, \emph {et~al.},\ }\href@noop {} {\bibfield  {journal} {\bibinfo  {journal} {APL Materials}\ }\textbf {\bibinfo {volume} {8}},\ \bibinfo {pages} {011109} (\bibinfo {year} {2020})}\BibitemShut {NoStop}%
\bibitem [{\citenamefont {Ohno}\ \emph {et~al.}(2022)\citenamefont {Ohno}, \citenamefont {Minami}, \citenamefont {Nakazawa}, \citenamefont {Sato}, \citenamefont {Kriener}, \citenamefont {Arita}, \citenamefont {Kawasaki},\ and\ \citenamefont {Uchida}}]{ohno2022maximizing}%
  \BibitemOpen
  \bibfield  {author} {\bibinfo {author} {\bibfnamefont {M.}~\bibnamefont {Ohno}}, \bibinfo {author} {\bibfnamefont {S.}~\bibnamefont {Minami}}, \bibinfo {author} {\bibfnamefont {Y.}~\bibnamefont {Nakazawa}}, \bibinfo {author} {\bibfnamefont {S.}~\bibnamefont {Sato}}, \bibinfo {author} {\bibfnamefont {M.}~\bibnamefont {Kriener}}, \bibinfo {author} {\bibfnamefont {R.}~\bibnamefont {Arita}}, \bibinfo {author} {\bibfnamefont {M.}~\bibnamefont {Kawasaki}},\ and\ \bibinfo {author} {\bibfnamefont {M.}~\bibnamefont {Uchida}},\ }\href@noop {} {\bibfield  {journal} {\bibinfo  {journal} {Physical Review B}\ }\textbf {\bibinfo {volume} {105}},\ \bibinfo {pages} {L201101} (\bibinfo {year} {2022})}\BibitemShut {NoStop}%
\bibitem [{\citenamefont {Nakamura}\ \emph {et~al.}(2024{\natexlab{b}})\citenamefont {Nakamura}, \citenamefont {Nishihaya}, \citenamefont {Ishizuka}, \citenamefont {Kriener}, \citenamefont {Ohno}, \citenamefont {Watanabe}, \citenamefont {Kawasaki},\ and\ \citenamefont {Uchida}}]{nakamura2024berry}%
  \BibitemOpen
  \bibfield  {author} {\bibinfo {author} {\bibfnamefont {A.}~\bibnamefont {Nakamura}}, \bibinfo {author} {\bibfnamefont {S.}~\bibnamefont {Nishihaya}}, \bibinfo {author} {\bibfnamefont {H.}~\bibnamefont {Ishizuka}}, \bibinfo {author} {\bibfnamefont {M.}~\bibnamefont {Kriener}}, \bibinfo {author} {\bibfnamefont {M.}~\bibnamefont {Ohno}}, \bibinfo {author} {\bibfnamefont {Y.}~\bibnamefont {Watanabe}}, \bibinfo {author} {\bibfnamefont {M.}~\bibnamefont {Kawasaki}},\ and\ \bibinfo {author} {\bibfnamefont {M.}~\bibnamefont {Uchida}},\ }\href@noop {} {\bibfield  {journal} {\bibinfo  {journal} {Physical Review B}\ }\textbf {\bibinfo {volume} {109}},\ \bibinfo {pages} {L121108} (\bibinfo {year} {2024}{\natexlab{b}})}\BibitemShut {NoStop}%
\bibitem [{\citenamefont {Sprague}\ \emph {et~al.}(2024)\citenamefont {Sprague}, \citenamefont {Regmi}, \citenamefont {Ghosh}, \citenamefont {Sakhya}, \citenamefont {Mondal}, \citenamefont {Bin~Elius}, \citenamefont {Valadez}, \citenamefont {Singh}, \citenamefont {Romanova}, \citenamefont {Kaczorowski} \emph {et~al.}}]{sprague2024observation}%
  \BibitemOpen
  \bibfield  {author} {\bibinfo {author} {\bibfnamefont {M.~X.}\ \bibnamefont {Sprague}}, \bibinfo {author} {\bibfnamefont {S.}~\bibnamefont {Regmi}}, \bibinfo {author} {\bibfnamefont {B.}~\bibnamefont {Ghosh}}, \bibinfo {author} {\bibfnamefont {A.~P.}\ \bibnamefont {Sakhya}}, \bibinfo {author} {\bibfnamefont {M.~I.}\ \bibnamefont {Mondal}}, \bibinfo {author} {\bibfnamefont {I.}~\bibnamefont {Bin~Elius}}, \bibinfo {author} {\bibfnamefont {N.}~\bibnamefont {Valadez}}, \bibinfo {author} {\bibfnamefont {B.}~\bibnamefont {Singh}}, \bibinfo {author} {\bibfnamefont {T.}~\bibnamefont {Romanova}}, \bibinfo {author} {\bibfnamefont {D.}~\bibnamefont {Kaczorowski}}, \emph {et~al.},\ }\href@noop {} {\bibfield  {journal} {\bibinfo  {journal} {Physical Review B}\ }\textbf {\bibinfo {volume} {110}},\ \bibinfo {pages} {045130} (\bibinfo {year} {2024})}\BibitemShut {NoStop}%
\bibitem [{\citenamefont {Weber}\ \emph {et~al.}(2006)\citenamefont {Weber}, \citenamefont {Cosceev}, \citenamefont {Drobnik}, \citenamefont {Faisst}, \citenamefont {Grube}, \citenamefont {Nateprov}, \citenamefont {Pfleiderer}, \citenamefont {Uhlarz},\ and\ \citenamefont {L{\"o}hneysen}}]{weber2006low}%
  \BibitemOpen
  \bibfield  {author} {\bibinfo {author} {\bibfnamefont {F.}~\bibnamefont {Weber}}, \bibinfo {author} {\bibfnamefont {A.}~\bibnamefont {Cosceev}}, \bibinfo {author} {\bibfnamefont {S.}~\bibnamefont {Drobnik}}, \bibinfo {author} {\bibfnamefont {A.}~\bibnamefont {Faisst}}, \bibinfo {author} {\bibfnamefont {K.}~\bibnamefont {Grube}}, \bibinfo {author} {\bibfnamefont {A.}~\bibnamefont {Nateprov}}, \bibinfo {author} {\bibfnamefont {C.}~\bibnamefont {Pfleiderer}}, \bibinfo {author} {\bibfnamefont {M.}~\bibnamefont {Uhlarz}},\ and\ \bibinfo {author} {\bibfnamefont {H.~v.}\ \bibnamefont {L{\"o}hneysen}},\ }\href@noop {} {\bibfield  {journal} {\bibinfo  {journal} {Physical Review B—Condensed Matter and Materials Physics}\ }\textbf {\bibinfo {volume} {73}},\ \bibinfo {pages} {014427} (\bibinfo {year} {2006})}\BibitemShut {NoStop}%
\bibitem [{\citenamefont {Zhang}\ \emph {et~al.}(2008)\citenamefont {Zhang}, \citenamefont {Zhao}, \citenamefont {Grin}, \citenamefont {Wang}, \citenamefont {Tang}, \citenamefont {Man}, \citenamefont {Chen},\ and\ \citenamefont {Yang}}]{zhang2008new}%
  \BibitemOpen
  \bibfield  {author} {\bibinfo {author} {\bibfnamefont {H.}~\bibnamefont {Zhang}}, \bibinfo {author} {\bibfnamefont {J.-T.}\ \bibnamefont {Zhao}}, \bibinfo {author} {\bibfnamefont {Y.}~\bibnamefont {Grin}}, \bibinfo {author} {\bibfnamefont {X.-J.}\ \bibnamefont {Wang}}, \bibinfo {author} {\bibfnamefont {M.-B.}\ \bibnamefont {Tang}}, \bibinfo {author} {\bibfnamefont {Z.-Y.}\ \bibnamefont {Man}}, \bibinfo {author} {\bibfnamefont {H.-H.}\ \bibnamefont {Chen}},\ and\ \bibinfo {author} {\bibfnamefont {X.-X.}\ \bibnamefont {Yang}},\ }\href@noop {} {\bibfield  {journal} {\bibinfo  {journal} {The Journal of Chemical Physics}\ }\textbf {\bibinfo {volume} {129}} (\bibinfo {year} {2008})}\BibitemShut {NoStop}%
\bibitem [{\citenamefont {Singh}\ \emph {et~al.}(2024)\citenamefont {Singh}, \citenamefont {Pavlosiuk}, \citenamefont {Dan}, \citenamefont {Kaczorowski},\ and\ \citenamefont {Wi{\'s}niewski}}]{singh2024large}%
  \BibitemOpen
  \bibfield  {author} {\bibinfo {author} {\bibfnamefont {K.}~\bibnamefont {Singh}}, \bibinfo {author} {\bibfnamefont {O.}~\bibnamefont {Pavlosiuk}}, \bibinfo {author} {\bibfnamefont {S.}~\bibnamefont {Dan}}, \bibinfo {author} {\bibfnamefont {D.}~\bibnamefont {Kaczorowski}},\ and\ \bibinfo {author} {\bibfnamefont {P.}~\bibnamefont {Wi{\'s}niewski}},\ }\href@noop {} {\bibfield  {journal} {\bibinfo  {journal} {Physical Review B}\ }\textbf {\bibinfo {volume} {109}},\ \bibinfo {pages} {125107} (\bibinfo {year} {2024})}\BibitemShut {NoStop}%
\bibitem [{\citenamefont {Nishihaya}\ \emph {et~al.}(2024)\citenamefont {Nishihaya}, \citenamefont {Nakamura}, \citenamefont {Ohno}, \citenamefont {Kriener}, \citenamefont {Watanabe}, \citenamefont {Kawasaki},\ and\ \citenamefont {Uchida}}]{nishihaya2024intrinsic}%
  \BibitemOpen
  \bibfield  {author} {\bibinfo {author} {\bibfnamefont {S.}~\bibnamefont {Nishihaya}}, \bibinfo {author} {\bibfnamefont {A.}~\bibnamefont {Nakamura}}, \bibinfo {author} {\bibfnamefont {M.}~\bibnamefont {Ohno}}, \bibinfo {author} {\bibfnamefont {M.}~\bibnamefont {Kriener}}, \bibinfo {author} {\bibfnamefont {Y.}~\bibnamefont {Watanabe}}, \bibinfo {author} {\bibfnamefont {M.}~\bibnamefont {Kawasaki}},\ and\ \bibinfo {author} {\bibfnamefont {M.}~\bibnamefont {Uchida}},\ }\href@noop {} {\bibfield  {journal} {\bibinfo  {journal} {Applied Physics Letters}\ }\textbf {\bibinfo {volume} {124}} (\bibinfo {year} {2024})}\BibitemShut {NoStop}%
\bibitem [{\citenamefont {Onoda}\ \emph {et~al.}(2008)\citenamefont {Onoda}, \citenamefont {Sugimoto},\ and\ \citenamefont {Nagaosa}}]{onoda2008quantum}%
  \BibitemOpen
  \bibfield  {author} {\bibinfo {author} {\bibfnamefont {S.}~\bibnamefont {Onoda}}, \bibinfo {author} {\bibfnamefont {N.}~\bibnamefont {Sugimoto}},\ and\ \bibinfo {author} {\bibfnamefont {N.}~\bibnamefont {Nagaosa}},\ }\href@noop {} {\bibfield  {journal} {\bibinfo  {journal} {Physical Review B—Condensed Matter and Materials Physics}\ }\textbf {\bibinfo {volume} {77}},\ \bibinfo {pages} {165103} (\bibinfo {year} {2008})}\BibitemShut {NoStop}%
\bibitem [{\citenamefont {Kurumaji}(2023)}]{kurumaji2023symmetry}%
  \BibitemOpen
  \bibfield  {author} {\bibinfo {author} {\bibfnamefont {T.}~\bibnamefont {Kurumaji}},\ }\href@noop {} {\bibfield  {journal} {\bibinfo  {journal} {Physical Review Research}\ }\textbf {\bibinfo {volume} {5}},\ \bibinfo {pages} {023138} (\bibinfo {year} {2023})}\BibitemShut {NoStop}%
\bibitem [{sup()}]{supp}%
  \BibitemOpen
  \href@noop {} {}\bibinfo {howpublished} {\url{URL_will_be_inserted_by_publisher}}\BibitemShut {NoStop}%
\bibitem [{\citenamefont {Thomson}(1857)}]{thomson1857xix}%
  \BibitemOpen
  \bibfield  {author} {\bibinfo {author} {\bibfnamefont {W.}~\bibnamefont {Thomson}},\ }\href@noop {} {\bibfield  {journal} {\bibinfo  {journal} {Proceedings of the Royal Society of London}\ }\textbf {\bibinfo {volume} {8}},\ \bibinfo {pages} {546} (\bibinfo {year} {1857})}\BibitemShut {NoStop}%
\bibitem [{\citenamefont {McGuire}\ and\ \citenamefont {Potter}(1975)}]{mcguire1975anisotropic}%
  \BibitemOpen
  \bibfield  {author} {\bibinfo {author} {\bibfnamefont {T.}~\bibnamefont {McGuire}}\ and\ \bibinfo {author} {\bibfnamefont {R.}~\bibnamefont {Potter}},\ }\href@noop {} {\bibfield  {journal} {\bibinfo  {journal} {IEEE Transactions on Magnetics}\ }\textbf {\bibinfo {volume} {11}},\ \bibinfo {pages} {1018} (\bibinfo {year} {1975})}\BibitemShut {NoStop}%
\bibitem [{\citenamefont {Burkov}(2017)}]{burkov2017giant}%
  \BibitemOpen
  \bibfield  {author} {\bibinfo {author} {\bibfnamefont {A.}~\bibnamefont {Burkov}},\ }\href@noop {} {\bibfield  {journal} {\bibinfo  {journal} {Physical Review B}\ }\textbf {\bibinfo {volume} {96}},\ \bibinfo {pages} {041110} (\bibinfo {year} {2017})}\BibitemShut {NoStop}%
\bibitem [{\citenamefont {Nandy}\ \emph {et~al.}(2017)\citenamefont {Nandy}, \citenamefont {Sharma}, \citenamefont {Taraphder},\ and\ \citenamefont {Tewari}}]{nandy2017chiral}%
  \BibitemOpen
  \bibfield  {author} {\bibinfo {author} {\bibfnamefont {S.}~\bibnamefont {Nandy}}, \bibinfo {author} {\bibfnamefont {G.}~\bibnamefont {Sharma}}, \bibinfo {author} {\bibfnamefont {A.}~\bibnamefont {Taraphder}},\ and\ \bibinfo {author} {\bibfnamefont {S.}~\bibnamefont {Tewari}},\ }\href@noop {} {\bibfield  {journal} {\bibinfo  {journal} {Physical Review Letters}\ }\textbf {\bibinfo {volume} {119}},\ \bibinfo {pages} {176804} (\bibinfo {year} {2017})}\BibitemShut {NoStop}%
\bibitem [{\citenamefont {Kumar}\ \emph {et~al.}(2018)\citenamefont {Kumar}, \citenamefont {Guin}, \citenamefont {Felser},\ and\ \citenamefont {Shekhar}}]{kumar2018planar}%
  \BibitemOpen
  \bibfield  {author} {\bibinfo {author} {\bibfnamefont {N.}~\bibnamefont {Kumar}}, \bibinfo {author} {\bibfnamefont {S.~N.}\ \bibnamefont {Guin}}, \bibinfo {author} {\bibfnamefont {C.}~\bibnamefont {Felser}},\ and\ \bibinfo {author} {\bibfnamefont {C.}~\bibnamefont {Shekhar}},\ }\href@noop {} {\bibfield  {journal} {\bibinfo  {journal} {Physical Review B}\ }\textbf {\bibinfo {volume} {98}},\ \bibinfo {pages} {041103} (\bibinfo {year} {2018})}\BibitemShut {NoStop}%
\bibitem [{\citenamefont {Burkov}\ and\ \citenamefont {Balents}(2011)}]{burkov2011weyl}%
  \BibitemOpen
  \bibfield  {author} {\bibinfo {author} {\bibfnamefont {A.}~\bibnamefont {Burkov}}\ and\ \bibinfo {author} {\bibfnamefont {L.}~\bibnamefont {Balents}},\ }\href@noop {} {\bibfield  {journal} {\bibinfo  {journal} {Physical Review Letters}\ }\textbf {\bibinfo {volume} {107}},\ \bibinfo {pages} {127205} (\bibinfo {year} {2011})}\BibitemShut {NoStop}%
\bibitem [{\citenamefont {Ito}\ and\ \citenamefont {Nomura}(2017)}]{ito2017anomalous}%
  \BibitemOpen
  \bibfield  {author} {\bibinfo {author} {\bibfnamefont {N.}~\bibnamefont {Ito}}\ and\ \bibinfo {author} {\bibfnamefont {K.}~\bibnamefont {Nomura}},\ }\href@noop {} {\bibfield  {journal} {\bibinfo  {journal} {journal of the Physical Society of Japan}\ }\textbf {\bibinfo {volume} {86}},\ \bibinfo {pages} {063703} (\bibinfo {year} {2017})}\BibitemShut {NoStop}%
\bibitem [{\citenamefont {Takahashi}\ \emph {et~al.}(2018)\citenamefont {Takahashi}, \citenamefont {Ishizuka}, \citenamefont {Murata}, \citenamefont {Wang}, \citenamefont {Tokura}, \citenamefont {Nagaosa},\ and\ \citenamefont {Kawasaki}}]{takahashi2018anomalous}%
  \BibitemOpen
  \bibfield  {author} {\bibinfo {author} {\bibfnamefont {K.~S.}\ \bibnamefont {Takahashi}}, \bibinfo {author} {\bibfnamefont {H.}~\bibnamefont {Ishizuka}}, \bibinfo {author} {\bibfnamefont {T.}~\bibnamefont {Murata}}, \bibinfo {author} {\bibfnamefont {Q.~Y.}\ \bibnamefont {Wang}}, \bibinfo {author} {\bibfnamefont {Y.}~\bibnamefont {Tokura}}, \bibinfo {author} {\bibfnamefont {N.}~\bibnamefont {Nagaosa}},\ and\ \bibinfo {author} {\bibfnamefont {M.}~\bibnamefont {Kawasaki}},\ }\href@noop {} {\bibfield  {journal} {\bibinfo  {journal} {Science Advances}\ }\textbf {\bibinfo {volume} {4}},\ \bibinfo {pages} {eaar7880} (\bibinfo {year} {2018})}\BibitemShut {NoStop}%
\bibitem [{\citenamefont {Soh}\ \emph {et~al.}(2018)\citenamefont {Soh}, \citenamefont {Donnerer}, \citenamefont {Hughes}, \citenamefont {Schierle}, \citenamefont {Weschke}, \citenamefont {Prabhakaran},\ and\ \citenamefont {Boothroyd}}]{soh2018magnetic}%
  \BibitemOpen
  \bibfield  {author} {\bibinfo {author} {\bibfnamefont {J.-R.}\ \bibnamefont {Soh}}, \bibinfo {author} {\bibfnamefont {C.}~\bibnamefont {Donnerer}}, \bibinfo {author} {\bibfnamefont {K.}~\bibnamefont {Hughes}}, \bibinfo {author} {\bibfnamefont {E.}~\bibnamefont {Schierle}}, \bibinfo {author} {\bibfnamefont {E.}~\bibnamefont {Weschke}}, \bibinfo {author} {\bibfnamefont {D.}~\bibnamefont {Prabhakaran}},\ and\ \bibinfo {author} {\bibfnamefont {A.}~\bibnamefont {Boothroyd}},\ }\href@noop {} {\bibfield  {journal} {\bibinfo  {journal} {Physical Review B}\ }\textbf {\bibinfo {volume} {98}},\ \bibinfo {pages} {064419} (\bibinfo {year} {2018})}\BibitemShut {NoStop}%
\bibitem [{\citenamefont {Murakami}\ \emph {et~al.}(2004)\citenamefont {Murakami}, \citenamefont {Nagosa},\ and\ \citenamefont {Zhang}}]{murakami20042}%
  \BibitemOpen
  \bibfield  {author} {\bibinfo {author} {\bibfnamefont {S.}~\bibnamefont {Murakami}}, \bibinfo {author} {\bibfnamefont {N.}~\bibnamefont {Nagosa}},\ and\ \bibinfo {author} {\bibfnamefont {S.-C.}\ \bibnamefont {Zhang}},\ }\href@noop {} {\bibfield  {journal} {\bibinfo  {journal} {Physical Review B—Condensed Matter and Materials Physics}\ }\textbf {\bibinfo {volume} {69}},\ \bibinfo {pages} {235206} (\bibinfo {year} {2004})}\BibitemShut {NoStop}%
\end{thebibliography}

\providecommand{\noopsort}[1]{}\providecommand{\singleletter}[1]{#1}%

\clearpage
\newpage

\clearpage
\newpage

\clearpage
\newpage

\clearpage
\newpage

\end{document}